\documentclass{pasj00}

\begin{document}
\SetRunningHead{Y. Aikawa et al.}{H$_2$CO in the Protoplanetary Disk around
LkCa 15}
\Received{2002/10/06}
\Accepted{2002/12/13}

\title{Interferometric Observations of Formaldehyde in the Protoplanetary Disk
around LkCa 15}

\author{Yuri \textsc{Aikawa},\altaffilmark{1}
        Munetake \textsc{Momose},\altaffilmark{2}
        Wing-Fai {\sc Thi},\altaffilmark{3}
        Gerd-Jan {\sc van Zadelhoff},\altaffilmark{3}\\
        Chunhua {\sc Qi},\altaffilmark{4}
        Geoffrey A. {\sc Blake},\altaffilmark{5}
        and
        Ewine F. {\sc van Dishoeck}\altaffilmark{3}}
\altaffiltext{1}{Department of Earth and Planetary Sciences, Kobe University,
Kobe 657-8501}
 \email{aikawa@kobe-u.ac.jp}
\altaffiltext{2}{Institute of Astrophysics and Planetary Sceinces,
Ibaraki University,\\
Bunkyo 2-1-1, Mito, Ibaraki 310-8512}
\altaffiltext{3}{Leiden Observatory, PO Box 9513, 2300 RA Leiden,
The Netherlands}
\altaffiltext{4}{Center for Astrophysics, 60 Garden Street, Cambridge,
MA 02138, USA}
\altaffiltext{5}{Division of Geological and Planetary Sciences,
California Institute of Technology,\\
MS 150-21, Pasadena, CA91125, USA}


%


\maketitle

\begin{abstract}
Emission from the $2_{12}$ -- $1_{11}$ line of H$_2$CO has been detected
and marginally resolved toward LkCa 15 by the Nobeyama Millimeter Array.
The column density of H$_2$CO is higher than that observed in DM Tau and
than predicted by theoretical models of disk chemistry; also, the
line-intensity profile is less centrally peaked than that for CO.
A similar behavior is observed in other organic gaseous molecules in
the LkCa 15 disk. 
\end{abstract}

\section{Introduction}
It is now well-established that more than 50\% of young T Tauri stars are
surrounded by a disk of circumstellar material.
These disks are important because they are the birth sites of planetary
systems. The spectral energy distributions indicate that disk masses are
in the range of $\sim 10^{-3}$ -- $10^{-1} M_{\odot}$, which is consistent
with the solar nebula model proposed for the origin of our own solar system
(e.g., Beckwith, Sargent 1996).
Interferometer observations in the dust continuum
and CO line emission have spatially resolved some of these disks
(e.g., Saito et al. 1995; Dutrey et al. 1996; Hogerheijde et al.\ 1997;
Guilloteau, Dutrey 1998; Mundy et al.\ 2000; Duvert et al. 2000),
which are found to have radii of
$\sim 100$ -- 800 AU. However, our quantitative understanding of the
physical properties of such disks, e.g. their radial and
vertical temperature and density structures, or their gas survival timescales,
is still poor. Another important question is the chemical evolution
of gas and dust as it is transported from the interstellar medium to
the interiors of circumstellar disks, and the impact of this
chemistry on the nature of  icy planetesimals, such as comets and
Kuiper Belt objects (van Dishoeck, Blake 1998; Ehrenfreund et al.\ 1997).

We are involved in a project to systematically investigate the physical
and chemical properties of circumstellar disks. Single-dish spectra of simple
molecules such as CO, HCO$^+$, CN, HCN, CS, and H$_2$CO have been obtained at
the IRAM 30 m telescope, JCMT and CSO (van Zadelhoff et al. 2001; Thi
2002, Thi et al. in preparation).
The ratios of high-$J$ and low-$J$ lines indicate that
the detected gaseous molecules reside in a low-temperature (e.g. $\sim 30$ K)
region, which corresponds to disk radii beyond $\gtrsim 100$ AU, assuming
the temperature distribution in typical disk models. In addition to
single-dish observations, interferometer observations of low-$J$ transition
lines of several molecular species have been performed by Duvert et al. (2000)
and Qi (2001).
Although interferometer observations of molecular lines in disks are
still rare because of the faintness of the objects, they provide an 
important probe of the physical and chemical gradients in the disks; the line
intensities depend on the density, temperature, and molecular abundance,
which can be affected by various chemical processes in each region
of the disk.

In this paper we report on observations of the H$_2$CO $2_{12}$ -- $1_{11}$
line at 140.84 GHz toward LkCa 15 using the Nobeyama Millimeter Array (NMA).
LkCa 15 is a relatively old ($\sim 1\times 10^7$ yr) solar-mass T Tauri star
located in the outer regions of the Taurus molecular cloud at $\sim 140$ pc.
Its disk mass is estimated to be $\sim 0.02$ -- $0.06 M_{\odot}$
from dust continuum observations (i.e. Thi 2002; Kitamura et al. 2002).
The formaldehyde molecule is of special interest, because its line ratios
are sensitive to the temperature and density (e.g. Mangum, Wootten 1993).
Single-dish observations of various H$_2$CO lines have been obtained by
Thi (2002) using the IRAM 30 m telescope and JCMT.
It is also one of the most complicated molecules detected so far in disks,
serves as a probe of grain-surface chemistry, and
can rapidly polymerize to create complex organic solids under appropriate
conditions (Schutte et al. 1993).
Observations of the transitions in the 1.3 mm atmospheric window using OVRO
and a comparison with the data presented here
will be reported in a forthcoming paper.

\section{Observations}

We observed the $2_{12}$ -- $1_{11}$ line of H$_2$CO toward LkCa 15 with
the NMA in 2000 -- 2001. The results presented here are based primarily
on 16 hr of array measurements using six antennas in the NMA
low-resolution (D) configuration, under clear-sky conditions. Similar
quality data were obtained over a 6-hr period in the high-resolution
(AB) configuration, but no line emission was detected. The beam sizes
of the D and AB configurations are about $4.''5$ and $1.''2$, respectively.

A dust continuum emission at $\lambda =2$ mm was detected 
with a good signal-to-noise ratio with the NMA SIS double sideband (DSB)
receivers. The zenith DSB system temperatures during the observations
were typically 200 K. For back ends, the Ultra Wide Band Correlator
(UWBC, Okumura et al. 2000) and a high-dispersion FX correlator were 
operated simultaneously. Phase-switching techniques were used to
separate the continuum visibility data for the lower ($ 128.840 \pm 
0.512$ GHz) and upper ($ 140.840 \pm 0.512$ GHz) sidebands from 
the UWBC. Spectral line visibility data collected by the FX 
correlator were obtained by subtracting the continuum level, which
was estimated by averaging the line-free channels. 

The response across the observed passband for each sideband was 
determined from 40-min observations of 3C 454.3. The gain calibrator
0446+112, used to determine the flux density scale, was observed every
20 min. The 2 mm flux density of 0446+112 was estimated to be
1.57 Jy at the time of these observations from a comparison with
the visibilities measured for Uranus (Griffin, Orton 1993), with an
overall uncertainty in the flux calibration of $\sim$10\%.
Final images were made using the calibrated visibilities within
the AIPS package developed at NRAO. 

\section{Results}
\subsection{Detection of H$_2$CO in the Disk of LkCa 15}

\begin{figure}
  \begin{center}
    \FigureFile(80mm,80mm){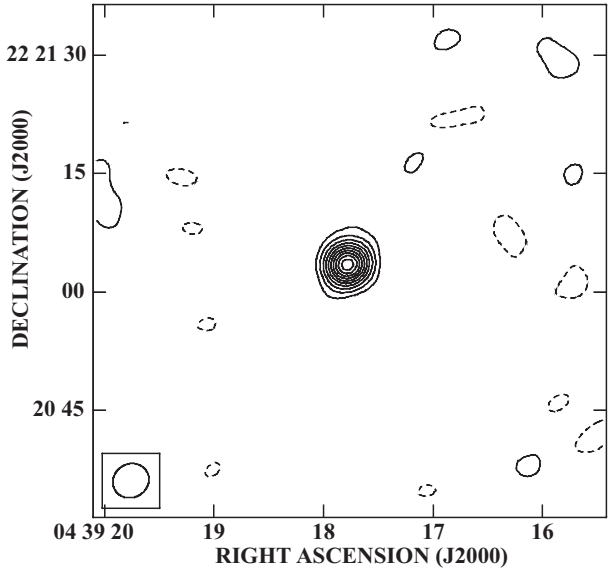}
  \end{center}
  \caption{Integrated intensity dust continuum map of the central $60''$
region towards LkCa 15.
The contour intervals are 2 $\sigma$, starting at $\pm 2$ $\sigma$ (1 $\sigma =
1.173$ mJy beam$^{-1}$). Negative contours are shown as dashed lines.}
\label{fig:cont}
\end{figure}

The continuum emission from the LkCa 15 disk was clearly detected;
an image using only the low-resolution configuration upper
sideband data is shown in figure \ref{fig:cont}.
The total dust flux at 141 GHz is 25 mJy and is unresolved, leading
to an upper limit to the dust disk radius of $\sim 550$ AU.

\begin{figure}
  \begin{center}
    \FigureFile(75mm,75mm){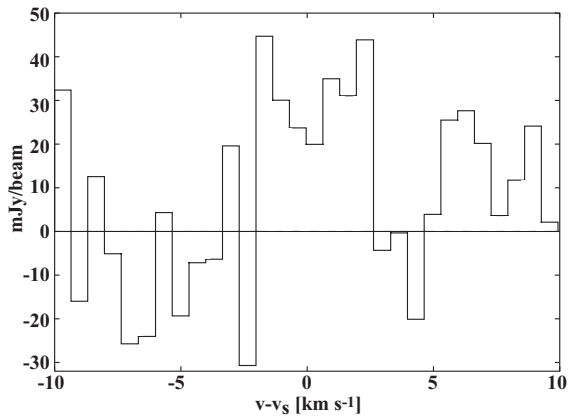}
  \end{center}
  \caption{Line spectrum of H$_2$CO $2_{12}$ -- $1_{11}$ in the central $10''$
region of the map. The stellar velocity $v_{\rm s}$ is
6 km s$^{-1}$.}\label{fig:spec}
\end{figure}

\begin{figure}
  \begin{center}
    \FigureFile(80mm,80mm){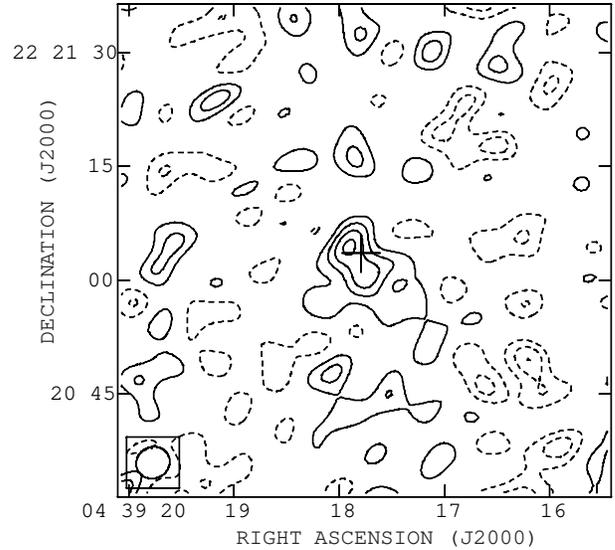}
  \end{center}
  \caption{Intensity map of H$_2$CO $2_{12}$ -- $1_{11}$ integrated from
$v-v_{\rm s}=-2$ to +2.6 km s$^{-1}$. The contour intervals are 1 $\sigma$,
starting at $\pm 1$ $\sigma$ (1 $\sigma=21$ mJy beam$^{-1}$). Negative
contours are shown as dashed lines. The cross indicates the peak position
of the dust continuum.}\label{fig:channel}
\end{figure}

The line spectrum for a 10$''$ box centered on the stellar position
is shown in figure \ref{fig:spec}. Emission was detected in the range
$\pm 2.5$ km s$^{-1}$ around the stellar velocity $v_{\rm s}=v_{\rm LSR} = 6$
km s$^{-1}$. Since the signal-to-noise ratio is not high enough to draw a
image with the velocity bin of $\sim 0.7$ km s$^{-1}$ adopted in figure
\ref{fig:spec},
we integrated the data from $v-v_{\rm s}= -2$ to $+2.6$ km s$^{-1}$ to obtain
figure \ref{fig:channel}.
The contour interval is 21 mJy beam$^{-1}$,
which corresponds to the 1 $\sigma$ r.m.s. noise
level. The position of the H$_2$CO emission coincides with that of the dust
continuum. It should also be noted that
previous CO observations show no foreground or background molecular gas
towards the object (Duvert et al. 2000). We therefore conclude that
the H$_2$CO emission arises from the disk around LkCa 15. The line
spectrum is consistent with the double-peaked shape characteristic
of Keplerian disk velocity fields, although we could not detect any
gradients in the channel maps with higher kinematic resolution 
(e.g. 1 km s$^{-1}$) because of the low signal-to-noise ratio.

The H$_2$CO disk is marginally resolved; it is more extended than the
continuum image, which is typical for molecular line images because of
the increased opacity as compared to that of the dust (Dutrey et al. 
1996). The H$_2$CO disk elongation from north-west to south-east is
caused by the disk inclination from the line of sight. Interferometer
maps of CN line emission show similar elongation along the same direction,
and CO observations reveal a velocity gradient that runs along this
major axis (Duvert et al. 2000; Qi 2001). The radius of the H$_2$CO disk,
measured at the 2-$\sigma$ level, is about 650 AU, which is similar
to the CO disk radius deduced from Plateau de Bure interferometer 
observations (Duvert et al. 2000).

The total flux density of the H$_2$CO line averaged over the 4.6 km s$^{-1}$
interval at the stellar velocity is about 180 mJy, leading to an
integrated line intensity of 820 mJy km s$^{-1}$. This is similar to
the value obtained from single-dish data at the IRAM 30 m telescope by
Thi (2002), which is about 664 mJy km s$^{-1}$,
as is the estimated line width
of $\sim 4$ km s$^{-1}$.  The interferometer
observations therefore trace the same gaseous component detected
by the IRAM 30 m telescope.

\subsection{Molecular Column Density}
The peak flux of the H$_2$CO line is about 95 mJy beam$^{-1}$, which
corresponds to an antenna temperature ($T_{\rm A}^*$) of 0.28 K for
a $4.''48$ $\times$ $4.''13$ synthesized beam.
The averaged antenna temperature of the line within the 2 $\sigma$
contour is about 0.16 K. Considering typical disk models,
because the gas density is higher than the critical density of the
line, $\sim 10^5$ cm$^{-3}$, the line is almost in local thermodynamic
equilibrium (LTE).
Since the disk image is marginally resolved, it is not likely that the
beam-filling factor is much smaller than unity. If the line is
optically thick, with the above two conditions, the antenna
temperature should be close to the kinetic temperature of the molecular gas.
The observed antenna temperature, however, is much lower than the kinetic
temperature of molecular gases in typical disk models, $\gtrsim 20$ K.
Hence, the molecular line is likely to be optically thin. In this subsection
we derive the molecular column density in the disk using LTE models.

The averaged antenna temperature, 0.16 K, leads to an averaged upper level
column density of $\sim 5.4 \times 10^{11}$ cm$^{-2}$ from the formula
$N_{\rm up}=8\pi k \nu^2 T_{\rm a} \triangle v/(hc^3 A_{\rm ul})$,
in which $A_{\rm ul} = 5.3 \times 10^{-5}$ s$^{-1}$ is the
Einstein $A$ coefficient (e.g. Goldsmith, Langer 1999). The total
column density of ortho H$_2$CO can be calculated from
$N_{\rm tot}=N_{\rm up} Q / [g_{\rm up} \exp (-E/kT)]$, where $Q$
is the rotational partition function, $g_{\rm up} =15$ is the
degeneracy of the upper state, and $E =21.9$ K is the upper
state energy. Theoretical models of disk chemistry predict 
gaseous organic molecules exist predominantly in regions where
$T\gtrsim 20$ K (section 4), while recent disk radiative transfer models
predict that the surface region of the disk can be heated to
temperatures of $\sim 50$ K by the central star (Chiang, Goldreich
1997; D'Alessio et al. 1998). For $T_{\rm ROT}$ values of 20 or 50 K,
the column density of ortho H$_2$CO is $5.4 \times 10^{12}$ cm$^{-2}$
or $1.1 \times 10^{13}$ cm$^{-2}$, respectively, in LTE. The
unknown ortho--para ratio is another uncertainty in estimating the total 
column density of H$_2$CO. In interstellar clouds the formaldehyde
ortho-para ratio ranges from 1.5 -- 3.0 (e.g. Mangum, Wootten 1993).
Hence, the total H$_2$CO column density averaged within the 2 $\sigma$ contour,
which corresponds to a region of about 970 AU $\times$ 1250 AU, is estimated
to be $7.2 \times 10^{12} - 1.9 \times 10^{13}$ cm$^{-2}$.

\section{Discussion}
DM Tau is another T Tauri star/disk system toward which H$_2$CO has
been detected (Dutrey et al. 1997). The estimated molecular column
density is $\sim 1 \times 10^{12}$ cm$^{-2}$, which is much smaller
than that in LkCa 15. Observations of other molecular lines also show
that the column densities of gaseous organic molecules are higher in
LkCa 15 than in DM Tau by a factor of $\sim 10$ (Qi 2001).

\begin{figure}
  \begin{center}
    \FigureFile(90mm,90mm){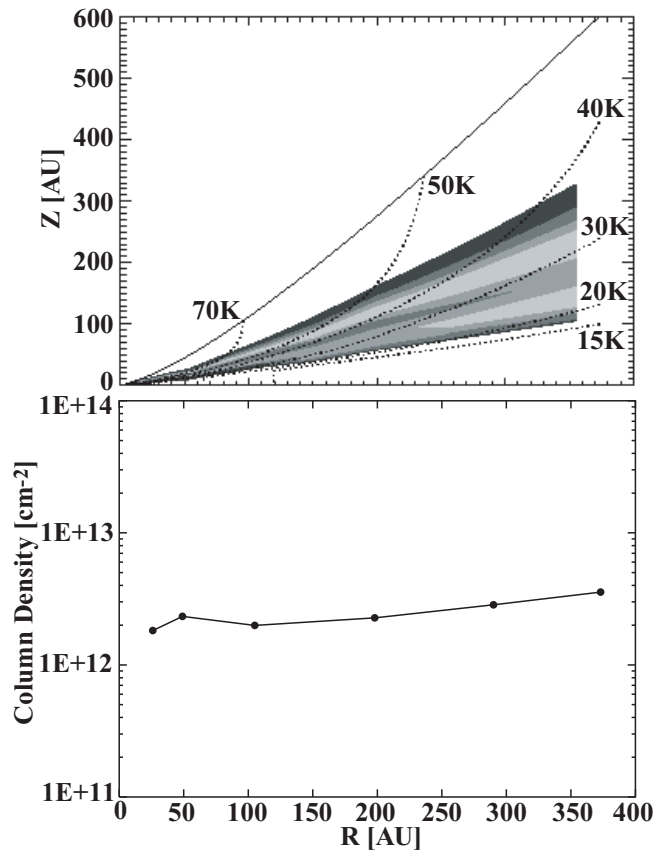}
  \end{center}
  \caption{Distributions of abundance and column density of H$_2$CO predicted
by van Zadelhoff et al. (2003). It is their model with Spectrum B;
including excess UV radiation from the central star with a spectrum similar
to that observed for TW Hya.
The gray scale in the upper panel shows regions
in which H$_2$CO abundance relative to hydrogen nuclei is $<
5\times 10^{-13}$ (white), $5\times 10^{-13}
-1\times 10^{-12}$ (the darkest gray), $1\times 10^{-12}- 1\times 10^{-11}$,
$1\times 10^{-11}-1\times 10^{-10}$ and $1\times 10^{-10}- 2.5\times 10^{-10}$
(the lightest gray). The line contours depict the temperature distribution.
The lower panel presents the radial distribution of the H$_2$CO column density
integrated in the vertical direction.}
\label{fig:model}
\end{figure}

The molecular abundances in the outer regions ($R\gtrsim 100$ AU)
of disks have been theoretically investigated by several groups
(Aikawa, Herbst 1999, 2001; Willacy, Langer 2000; Aikawa 
et al. 2002; van Zadelhoff et al. 2003). In all such models,
the chemistry is intimately tied to the physical state of the
gas and the properties of the disk. For example, the upper panel of figure
\ref{fig:model} shows the vertical and radial distribution of H$_2$CO in the
model by van Zadelhoff et al. (2003).
They adopted the disk model of D'Alessio et al. (1999), and solved the
ultra-violet (UV) radiation transfer and chemical reaction network,
which includes gas-phase chemistry,
freeze-out and evaporation, but no active surface chemistry.
Near the disk surface, molecules
are dissociated by UV radiation from the interstellar
field and central star. In the disk midplane organic molecules
are heavily depleted onto grains, since the densities are very
high and the temperature is lower than 20 K, the sublimation
temperature of the dominant carbon reservoir, CO. Thus, in the
outer reaches of circumstellar disks, gaseous molecules exist in
high abundance only in an intermediate height layer that is heated
by the infrared radiation from the surface layer yet shielded
from the harsh UV radiation field. 
Specifically, formaldehyde is found to have a maximum gas-phase
abundance in gas at A$_V \sim 0.4$ -- 10 mag from the disk surface.

Model comparisons with observations require knowledge of the
molecular column density versus distance from the central star, and the lower
panel of figure \ref{fig:model} shows the H$_2$CO column density calculated
by van Zadelhoff et al. (2003). Their result is in reasonable agreement
with observations of DM Tau, but smaller than that needed to
reproduce the LkCa 15 observations. Clearly, there must be some
difference in the physical structure or evolutionary paths of DM Tau
and LkCa 15 which causes the chemical variation seen in these two
objects. Interestingly, the models predict that the molecular
column densities do not significantly depend on either the
disk mass or age; that the mass of the intermediate molecular layer
does not depend on the total disk mass as long as the disk is as thick as
$A_V \gtrsim 20$ mag, and that the chemical timescale
in the intermediate layer is relatively short, $\sim 10^4$ -- $10^5$ yr.
On the other hand, X-rays from the central star may cause variations
in the molecular abundances, since the X-ray-induced UV destruction of
CO can supply the additional carbon needed to form other species.
Neither DM Tau nor LkCa 15 has been detected by ROSAT X-ray surveys. However,
considering the high time variability of X-rays from T Tauri stars,
reobservation of these objects with Chandra would be highly desirable. 
Dust coagulation and sedimentation could also cause variations of the
molecular abundances since such processes would significantly change
the temperature distribution and UV radiation field within the disks.
Indeed, recent studies on the spectral energy distribution indicate the
coagulation and settling of dust in
several T Tauri and Herbig Ae/Be disks, including LkCa 15 (Chiang et al. 2001;
D'Alessio et al. 2001). Finally, active grain surface chemistry
with evaporation may enhance the H$_2$CO abundance, and may cause variations
of the molecular abundances among objects, if they have different temperature
or grain size.

The spatial distribution of molecules is another interesting issue
concerning the chemical and physical structure of protoplanetary disks.
Theoretical models predict that the H$_2$CO column density is fairly
insensitive to the total H$_2$ column density, and thus to the disk
radius. On the other hand, CO is one of the most volatile species
after molecular hydrogen, with a sublimation temperature of only 
$\sim 20$ K, which is reached in the midplane at $R\sim 100$ AU. This
causes a steep gradient in the radial distribution of the CO column density.
In our observations, the H$_2$CO line intensity is less centrally
peaked than that of CO, which was observed by Duvert et al. (2000). 
However, we cannot exclude the possibility that the flat distribution of
the H$_2$CO emission intensity is caused by the low spatial resolution and/or
low signal-to-noise ratio of our observations; the central region of the
Keplerian disk corresponds to the wing component of the line spectrum, which
is more easily affected by noise than velocities corresponding to
the outer regions of the disk.

In order to test these possibilities, we observed the LkCa 15 disk with
the high-resolution ($1.''2$) configuration of the NMA for 4 nights.
Unfortunately, clear-sky conditions were present for only 6 hr,
leading to a noise level of 82 mJy beam$^{-1}$ or 3.1 K.
Since the peak antenna temperature of 0.28 K in the D
configuration corresponds to 3.6 K in a $1.''2$ beam, these high-resolution
observations do not add any significant constraints on the spatial distribution
of H$_2$CO.
Some quantitative constraints are given as follows.
If the temperature of the H$_2$CO at $R=150$ AU is similar to that
at $R=600$ AU, the ratio of the total H$_2$CO
column density at these radii should be proportional to that of brightness
(i.e. mean antenna temperature), and is less than $3.1/0.16 \sim 19$. If the
molecular layer in the inner radius has a higher temperature than that at
the outer radius, we obtain a weaker constraint. Assuming temperatures of
50 K and 20 K at $R=150$ AU and 600 AU, respectively, the column density
ratio is less than 40.
Observations with higher sensitivity and higher spatial resolution are
therefore critical for an improved understanding of the molecular
distributions and chemical processes within protoplanetary disks. 


\end{document}